\documentclass{emulateapj}
\usepackage{apjfonts}

\slugcomment{Submitted 2012 July 9; accepted 2012 September 4}

\shorttitle{Circumbinary disk dispersal}
\shortauthors{Alexander}

\begin{document}

% Macros
\newcommand{\Msunyr}{M$_{\odot}$yr$^{-1}$}
\newcommand{\Mjup}{M$_{\mathrm {Jup}}$}
\newcommand{\Msun}{M$_{\odot}$}

\title{The dispersal of protoplanetary disks around binary stars}

\author{Richard Alexander}
\affil{Department of Physics \& Astronomy, University of Leicester, Leicester, LE1 7RH, UK}
\email{richard.alexander@leicester.ac.uk}

%%%%%%%%%%%%%%%%%%%%%%%%%%%%

\begin{abstract}
\noindent I present models of disk evolution around young binary stars.  I show that the primary factor in determining circumbinary disk lifetimes is the rate of disk photoevaporation.  I also find that photoevaporative clearing leaves a signature on the distribution of circumbinary disk lifetimes, with a sharp increase in disk lifetimes for binary separations $a \lesssim $0.3--1AU.  Observations of young binary stars can therefore be used to test models of disk evolution, and I show that current data set a strong upper limit to the rate of on-going photoevaporation ($< 10^{-9}$\Msunyr).  Finally I discuss the implications of these results for planet formation, and suggest that circumbinary planets around close ($a \lesssim 1$AU) binaries should be relatively common.
\end{abstract}


%%%%%%%%%%%%%%%%%%%%%%%%%%%%

\section{Introduction}\label{sec:intro}
It has long been recognised that most Sun-like stars form in binary or multiple systems \citep[e.g.,][]{dm91}.  Circumbinary disks are a natural consequence of binary star formation \citep[e.g.,][]{monin_ppv}, and these young disks are potentially the sites of planet formation.  Recently the {\it Kepler} mission has discovered a number of circumbinary planets \citep{doyle11,welsh12,orosz12a,orosz12b}, the existence of which poses interesting challenges for planet formation theories.  Moreover, new high-resolution techniques have now begun to build a useful census of disks around, young, $\sim$AU-separation binary stars \citep[e.g.,][]{kraus12,harris12}.  These observations suggest that planets form readily in circumbinary disks, but little is known about the physical conditions in these complex young systems.

In this {\it Letter} I present a simple one-dimensional (1-D) model for the evolution of protoplanetary disks around close ($\lesssim 10$AU) binary stars.  The formation and early evolution of such disks is dominated by gravitational instabilities and magnetically-driven outflows \citep[e.g.,][]{durisen11,ks11}, while later evolution and final disk dispersal is driven by the competition between viscous accretion and photoevaporation \citep[e.g.,][]{cc01,acp06b}.  However, the tidal torque from a binary inhibits disk accretion and modifies the evolution substantially from the single-star case.  I find that the rate of disk photoevaporation plays a critical role in determining how young circumbinary disks evolve, and that photoevaporative clearing leaves a characteristic signature on the distribution of disk lifetimes.  I also show how observations of circumbinary disks can be used to inform our understanding of protoplanetary disk clearing, and discuss the consequences of these results for planet formation around binary stars.

%%%%%%%%%%%%%%%%%%%%%%%%%%%%
\section{Model}\label{sec:model}
The evolution of an accretion disk around a binary is described by \citep[e.g.,][]{lp86}
\begin{equation}\label{eq:1ddiff}
\frac{\partial \Sigma}{\partial t} = \frac{1}{R}\frac{\partial}{\partial R}\left[ 3R^{1/2} \frac{\partial}{\partial R}\left(\nu \Sigma R^{1/2}\right) - \frac{2 \Lambda \Sigma R^{3/2}}{(GM_{\mathrm {tot}})^{1/2}}\right] - \dot{\Sigma}_{\mathrm {w}}(R,t) \, .
\end{equation}
Here $\Sigma(R,t)$ is the disk surface density, $t$ is time, $R$ is the cylindrical radius (relative to the centre of mass), $\nu$ is the kinematic viscosity, $M_{\mathrm {tot}}$ is the total stellar mass (i.e., $M_{\mathrm {tot}} = M_1 + M_2$, where $M_1$ \& $M_2$ are the masses of the individual stars), and $\dot{\Sigma}_{\mathrm {w}}(R,t)$ is the mass-loss due to photoevaporation.  The binary orbit is assumed to be circular and co-planar with the disk.  For a binary of mass ratio $q = M_2/M_1$ and semi-major axis $a$, the tidal torque from the binary, $\Lambda(R,a)$, is approximated by\footnote{Strictly this expression for $\Lambda$ applies only for $R>a$, but I assume $\Sigma = 0$ interior to the binary orbit so the case $R < a$ is not relevant here.} \citep[e.g.,][]{armitage02}
\begin{equation}\label{eq:torque}
\Lambda(R,a) = \frac{q^2 GM_{\mathrm {tot}}}{2R} \left(\frac{a}{\Delta_{\mathrm p}}\right)^4 \, ,
\end{equation}
where
\begin{equation}
\Delta_{\mathrm p} = \textrm{max}(H,|R-a|)
\end{equation}
and $H$ is the disk scale-height.  The back-reaction of this torque causes the binary orbit to shrink, but this depends sensitively on the gas dynamics inside the inner edge of the disk (which cannot be accurately modeled in 1-D).  Moreover, the bulk of this evolution happens at early times, while here our primary interest is the late stages of the disk evolution, so for simplicity I fix the binary separation $a$ to be constant throughout.  Test calculations show that this approximation does not influence the results significantly.

Accretion in protoplanetary disks is thought to be driven by magnetohydrodynamic turbulence \citep[e.g.,][]{balbus11}.  Here this is approximated by an alpha-prescription for the disk viscosity $\nu(R) = \alpha \Omega H^2$, where $\alpha=0.01$ is the \citet{ss73} viscosity parameter and $\Omega(R) = \sqrt{GM_{\mathrm {tot}}/R^3}$.  $\alpha$ here essentially represents a time and space average of the efficiency of angular momentum transport in the disk, and the choice $\alpha=0.01$ is consistent with observations of protoplanetary disk accretion \citep[e.g.,][]{hcga98}.  I choose a power-law form for the disk scale-height $H \propto R^{5/4}$, which results in a linear viscosity law $\nu \propto R$.  The power-law is normalized by setting $H/R=0.05$ at $R=1$AU.

It is now well established that photoevaporation by high-energy photons dominates protoplanetary disk clearing at late times \citep[e.g.,][]{ps09,pascucci11}.  This radiation heats the disk surface, and beyond some critical radius the heated gas is unbound and flows as a wind.  When the disk is optically thick to the high-energy radiation this mass-loss is concentrated around the critical radius $R_{\mathrm c}\simeq 0.2 GM_{\mathrm {tot}}/c_{\mathrm s}^2$, where $c_{\mathrm s}$ is the sound speed of the heated disk atmosphere.  However, if there is an optically thin inner cavity the inner disk edge can be photoevaporated directly, and the mass-loss is instead concentrated close to the disk edge.  Around single stars this occurs only at late times, when the wind is able to overcome disk accretion.  Here, however, the torque from the binary can clear the inner disk, and change the qualitative behaviour of the photoevaporative wind. 

Unfortunately it is not yet clear whether X-ray \citep[e.g.,][]{owen10} or UV irradiation \citep[e.g.,][]{acp06b} drives the photoevaporative wind.  However, the mass-loss profile is similar in both cases, so for simplicity I adopt a parametrized form for the photoevaporative mass-loss term (motivated by analytical models, e.g., \citealt{holl94,acp06a}).  When the inner disk is optically thick 
\begin{equation}
\dot{\Sigma}(R) = \frac{\dot{M}_{\mathrm {thick}}}{4\pi R_{\mathrm c}^2} \left(\frac{R}{R_{\mathrm c}}\right)^{-5/2} \quad , \quad R \ge R_{\mathrm c} \, ,
\end{equation}
and when the inner disk is optically thin (i.e., the disk inner edge lies at $R_{\mathrm {in}} > R_{\mathrm c}$), we instead have
\begin{equation}
\dot{\Sigma}(R) = \frac{\dot{M}_{\mathrm {thin}}}{4\pi R_{\mathrm {in}}^2} \left(\frac{R}{R_{\mathrm {in}}}\right)^{-5/2} \left(\frac{R}{2 R_{\mathrm c}}\right)^{1/2} \quad , \quad R \ge R_{\mathrm {in}} \, ,
\end{equation}
With this form the mass-loss per unit area peaks at $R_{\mathrm c}$ (in the optically thick case), and $\dot{M}_{\mathrm {thick}}$ and $\dot{M}_{\mathrm {thin}}$ are the integrated mass-loss rates (normalised to $R_{\mathrm {in}} = 2R_{\mathrm c}$ in the optically thin case).  The second form for $\dot{\Sigma}(R)$ is used when the surface density interior to $R_{\mathrm c}$ falls below the critical value $\Sigma_{\mathrm {thick}}$, and this parametrization successfully mimics the behavior of more sophisticated models.  I define two wind models: one with a low wind rate and small critical radius, and a second with a higher wind rate and a larger critical radius.  The first represents photoevaporation by ionizing (EUV) photons \citep{font04,acp06a}, with $R_{\mathrm c} = 1.4$AU, $\dot{M}_{\mathrm {thick}} = 1.6\times10^{-10}$\Msunyr, $\dot{M}_{\mathrm {thin}} = 1.1\times10^{-9}$\Msunyr and $\Sigma_{\mathrm {thick}} = 10^{-5}$g\,cm$^{-2}$.  The second represents photoevaporation by X-rays \citep{owen10,owen11}, with $R_{\mathrm c} = 5.0$AU, $\dot{M}_{\mathrm {thick}} = \dot{M}_{\mathrm {thin}} = 1.0\times10^{-8}$\Msunyr, and $\Sigma_{\mathrm {thick}} = 10^{-2}$g\,cm$^{-2}$.  These two wind models are henceforth referred to as the ``weak'' and ``strong'' photoevaporative winds, respectively.

The form of the torque function $\Lambda$ does not allow any gas to accrete on to the binary, and in the absence of mass-loss the model therefore describes a decretion disk \citep{pringle91}.  In practice, however, we expect gas to accrete from the disk on to the binary via tidal streams.  This accretion flow is variable and is modulated by the binary orbit, but the average accretion rate on to the binary can be as much as $\sim10$\% of the steady-state disk accretion rate \citep[e.g.,][]{mm08}.  I therefore allow gas to accrete from the inner disk edge on to the binary at a fraction $\epsilon$ of the disk accretion rate (computed as $3 \pi \nu \Sigma$ at $R=3 R_{\mathrm {in}}$).  These accretion streams may be optically thick to the radiation that drives photoevaporation, but detailed modeling of how the streams ``shield'' the wind is beyond the scope of this initial investigation.  Instead I define an average surface density $\Sigma_{\mathrm {streams}} = 2 \dot{M}_{\mathrm {in}}/\Omega_{\mathrm b} R_{\mathrm {in}}^2$, 
where $\dot{M}_{\mathrm {in}}$ is the accretion rate from the inner disk edge and $\Omega_{\mathrm b}$ is the orbital frequency of the binary, and switch wind profiles only when $\Sigma_{\mathrm {streams}} < \Sigma_{\mathrm {thick}}$.

Here we are primarily interested in the late-time evolution of the disk, so I adopt a simplified set of initial conditions.  The initial disk mass is taken to be $M_{\mathrm d} = 10^{-1.5}$\Msun, and the initial surface density profile is assumed to be an exponentially-truncated power-law \citep[e.g.,][]{lbp74}
\begin{equation}
\Sigma(R) = \frac{M_{\mathrm d}}{2\pi R_0 R} \exp \left(-\frac{R}{R_0}\right) \quad , \quad R \ge 5a
\end{equation}
The inner disk edge is initially a step-function at $5a$, but rapidly relaxes to a self-consistent profile.  The scaling radius $R_0$ determines the initial disk size and is set, arbitrarily, to be $R_0 = 15a$ (i.e., three times the inner edge radius).  This implicitly assumes that the disk angular momentum scales with that of the binary; this assumption is not well justified, but in the absence of a first-principles model for binary formation this is the most sensible way to proceed.

This set of disk models is therefore described by four parameters: the binary separation $a$ and mass ratio $q$, the photoevaporation model (strong or weak), and the accretion efficiency $\epsilon$.  I have run grids of models with values of $\log_{10}(a) = -1.0$, $-0.99$, $-0.98$\ldots1.5 (i.e., $a = 0.1$--31.6AU).  The standard model grid uses an equal mass binary ($q = 1$) and allows gas to accrete on to the binary with $\epsilon=0.1$.  Variant models were also run with $q = 0.3$, $\epsilon = 0$ and $\epsilon= 0.01$, as well as a reference set of single-star models (which have $\Lambda =0$ and the same $R_0$ as the standard models).  All of the model grids were run with both the strong and weak photoevaporative winds.  The models were integrated forwards in time until the circumbinary disk was cleared: this is defined to be the point at which the disk inner edge $R_{\mathrm {in}} > 25$AU and $R_{\mathrm {in}} > 10a$.  Operationally, I solve Equation \ref{eq:1ddiff} using a standard first-order explicit scheme on an $R^{1/2}$-spaced grid \citep[e.g.,][]{pvw86}, using 1000 cells to span the range $[0.09\mathrm {AU},2500\mathrm{AU}]$.

%%%%%%%%%%%%%%%%%%%%%%

\section{Results}\label{sec:results}
\begin{figure}
\centering
\includegraphics[angle=270,width=\hsize]{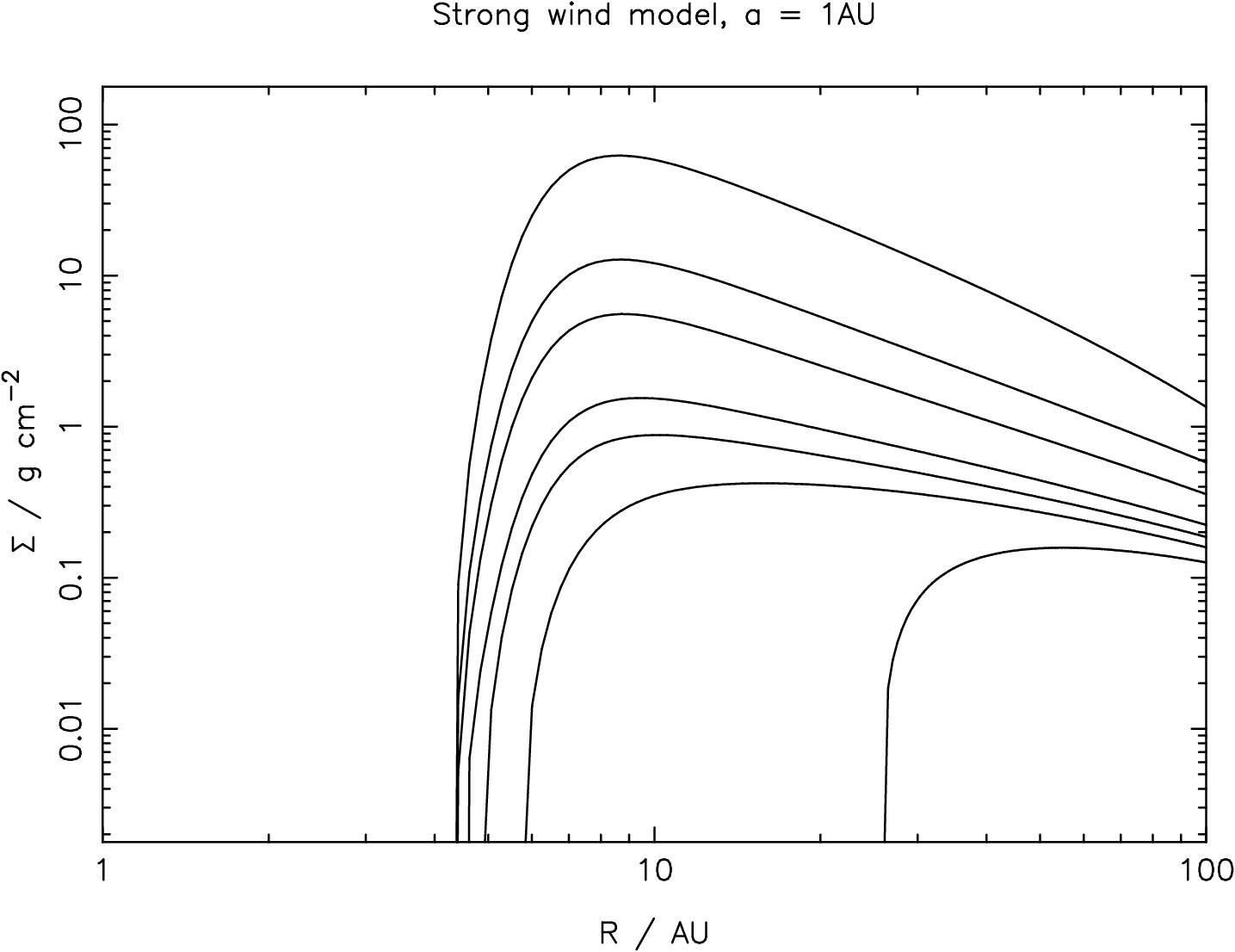}

\vspace*{5pt}

\includegraphics[angle=270,width=\hsize]{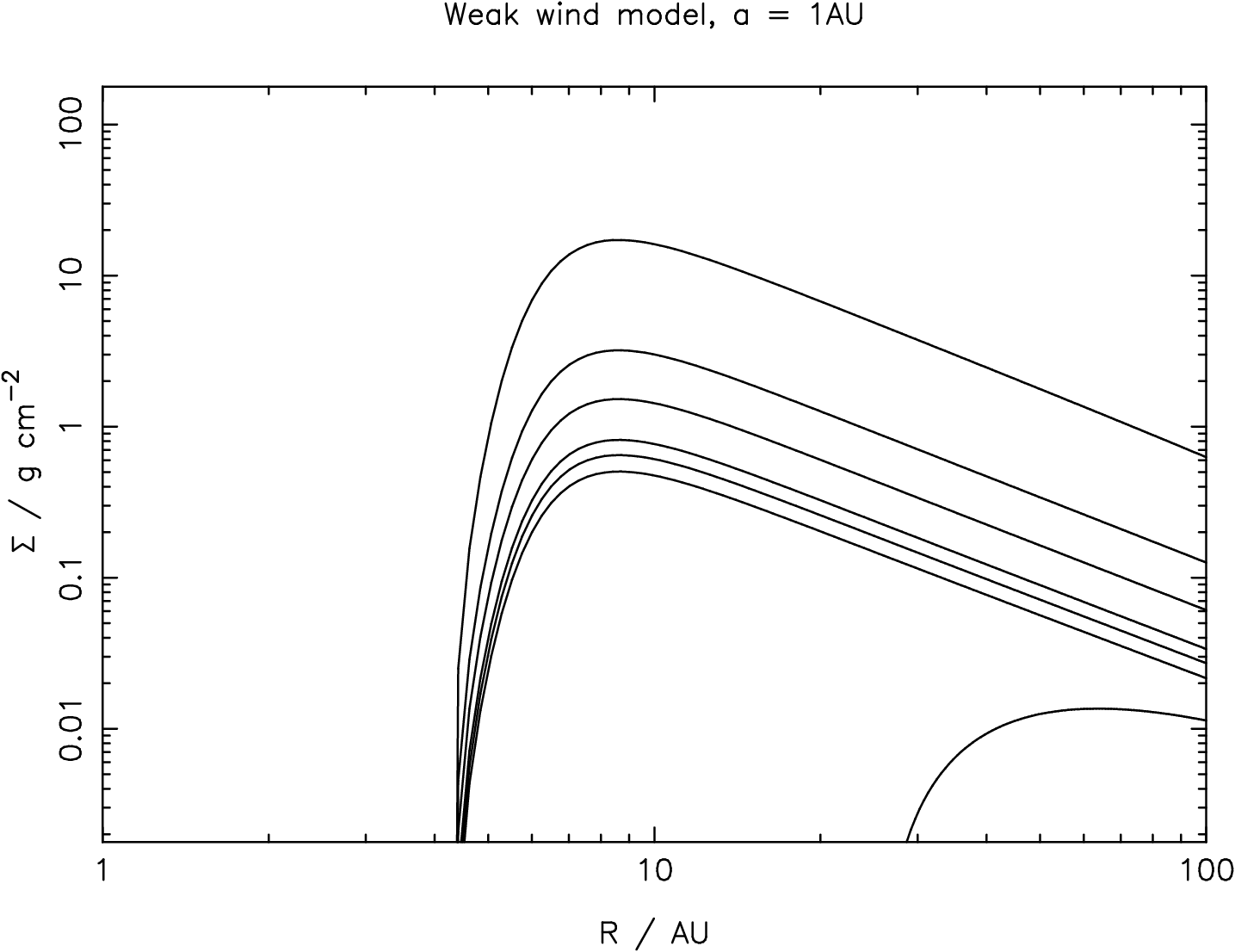}
\caption{Surface density evolution of the standard models with $a = 1$AU.  The upper panel shows the strong wind case (typical of X-ray photoevaporation), with $\Sigma(R)$ plotted at 0.1, 0.5, 0.7, 0.85, 0.9, 0.95 \& $1.0 t_{\mathrm {disk}}$.  (Here the disk lifetime $t_{\mathrm {disk}} =1.68$Myr).  In this case the majority of the disk is removed by the photoevaporative wind, and the disk is cleared rapidly once the surface density drops below $\simeq 1$g\,cm$^{-2}$.  The lower panel shows the weak wind case (typical of EUV photoevaporation), with $\Sigma(R)$ plotted at the same fractions of the disk lifetime ($t_{\mathrm {disk}} =13.74$Myr).  Here accretion dominates the evolution, and the wind only clears the disk once it has undergone substantial viscous evolution.}\label{fig:sigma_t}
\end{figure}
Figure \ref{fig:sigma_t} shows the evolution of the disk surface density in a characteristic model (the standard model with $a = 1$AU) for both the strong and weak wind cases.  In both cases the tidal torque from the binary slows disk evolution, increasing the lifetime by a factor of $\simeq2.5$--3 relative to an otherwise identical single-star disk.  The most obvious difference between the two models is that the disk lifetime $t_{\mathrm {disk}}$ is almost an order of magnitude shorter in the strong wind case (1.68Myr versus 13.74Myr; the reference single-star disk lifetimes are 0.57Myr and 5.51Myr, respectively).  Note, however, that the absolute values of the disk lifetime are not significant, as $t_{\mathrm {disk}}$ depends strongly on the initial conditions (primarily the initial disk mass).  It is more instructive to consider the evolution of the models as a function of the normalised time $t/t_{\mathrm {disk}}$, as this highlights the relative importance of the competing physical processes and eliminates most of the artefacts introduced by the choice of initial conditions.

If we compare these two models in this manner we see significant differences in their evolution.  In both cases the binary separation (1AU) is sufficiently small that the disk is optically thick interior to $R_{\mathrm c}$, and the initial accretion rate on to the binary is $3\times10^{-8}$\Msunyr.  This is much larger than the photoevaporation rate (henceforth $\dot{M}_{\mathrm w}$) in the weak wind case, but comparable to the wind rate in the strong wind case.  Consequently the two models follow qualitatively different evolutionary sequences.  In the strong wind case the photoevaporative wind almost immediately overwhelms the accretion flow.  The inner edge of the disk remains at $\simeq 4.2$AU for much of the disk lifetime, but is progressively eroded by the wind.  Over the disk lifetime most ($\simeq 70$\%) of the disk mass is removed by the photoevporative wind, with only $\simeq 30$\% accreting on to the binary.  The high mass loss rate also has a strong impact on the radial profile of the disk: the surface density never reaches the power-law profile characteristic of accretion- or decretion-dominated disks.  Eventually the wind prevails completely, and the disk is then rapidly cleared from the inside out.  This final clearing retains the ``two-time-scale'' behavior characteristic of photoevaporative disk dispersal \citep{cc01,acp06b}, but with the important caveat that accretion and angular momentum transport play only a minor role in the disk's evolution.

\begin{figure}
\centering
\includegraphics[angle=270,width=\hsize]{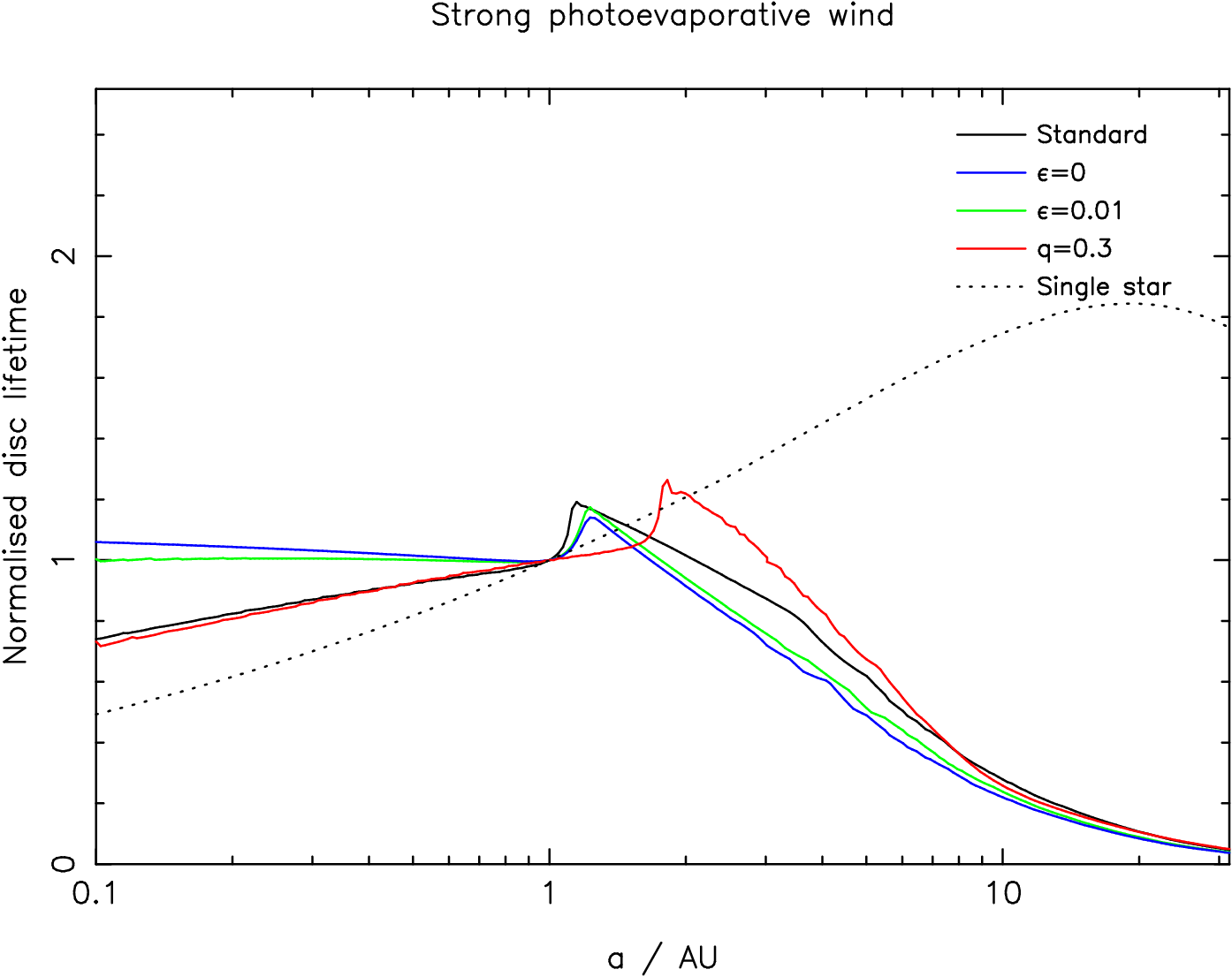}

\vspace*{5pt}

\includegraphics[angle=270,width=\hsize]{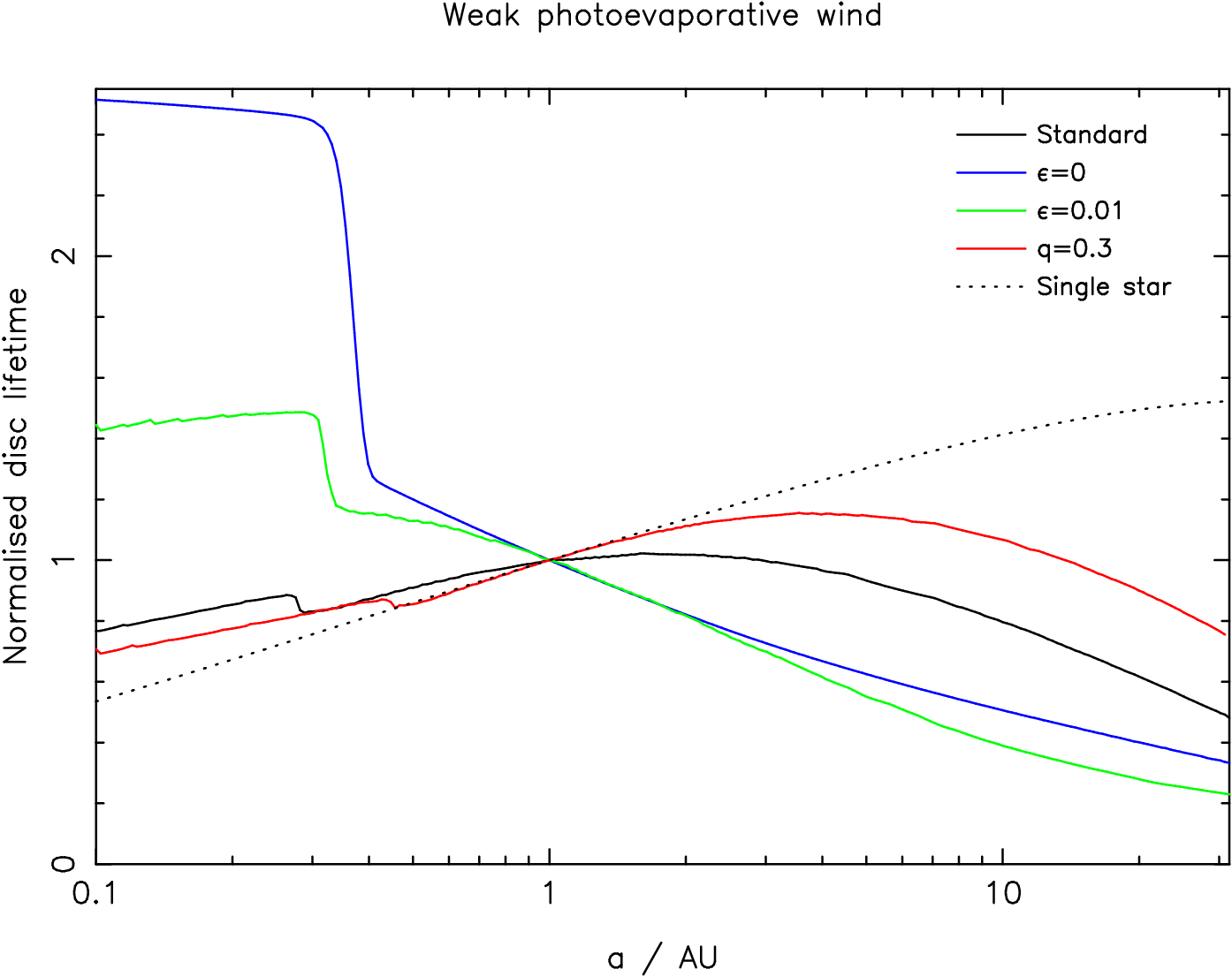}
\caption{Disk lifetimes $t_{\mathrm {disk}}$ plotted as a function of binary separation $a$.  For clarity, all the curves are normalised to the lifetime at $a = 1$AU.  The single-star models, plotted for reference, have the same viscous time-scales as the standard model set.  The upper panel shows the strong wind models; the lower panel the weak wind models.  The sharp increase in disk lifetimes for $a < 0.25R_{\mathrm c}$ is caused by the ``switch''  to direct photoevaporation of the disk once the binary has cleared a sufficiently large inner cavity.}\label{fig:lifetimes}
\end{figure}

By contrast, in the weak wind case the initial accretion rate exceeds the wind rate by a factor $\gtrsim 100$, and the disk's evolution is governed primarily by the viscosity.  At $R \gtrsim 10$AU the disk is close to the $\Sigma \propto R^{-3/2}$ power-law of a decretion disk and, as in the single-star case, only once the disk has undergone substantial viscous evolution does the photoevaporative wind trigger disk clearing.  The inner edge of the disk remains at a fixed position throughout, and although the final clearing again shows the characteristic two-time-scale behaviour, here the majority of the disk has been accreted on to the binary, with only a small fraction (7\%) removed by photoevaporation.

Further insight can be gained by looking at Figure \ref{fig:lifetimes}, which shows how the disk lifetime varies as a function of binary separation $a$ in the different models.  The most prominent features are a sharp increase in the disk lifetime at small separations ($a \lesssim 0.25R_{\mathrm c}$), and a progressive decrease in disk lifetimes at large separations ($a \gtrsim 5$--10AU).  The increase in disk lifetimes at small $a$ occurs because the photoevaporative wind has a characteristic radius: when the binary clears a cavity in the disk that is larger than the critical radius $R_{\mathrm c}$, the wind is driven by direct irradiation of the inner disk edge.  This increases the efficiency of the wind, and clears the disk more rapidly.  The increase in $t_{\mathrm {disk}}$ for small $a$ is much larger when accretion on to the binary is suppressed, and the critical value of $a$ also increases for a lower mass ratio $q$.  This effect is also more pronounced in the weak wind models.  This is primarily due to the more prominent role of accretion in the weak wind models, but is also partly because direct irradiation increases the weak wind rate\footnote{This is due to the increased efficiency of radiative transfer in the EUV wind once the inner disk is removed \citep{acp06a}.} by a factor of $\sim 10$.  By contrast, in the strong wind case photoevaporation dominates the evolution throughout and $t_{\mathrm {disk}} \sim M_{\mathrm d}/\dot{M}_{\mathrm w}$, so we see only small variations in disk lifetime at small $a$.

At large $a$ ($\gtrsim 5$--10AU) photoevaporation also dominates the disk evolution: accretion in these large disks is not efficient (as the viscous time-scale is long), and ${M}_{\mathrm w}$ increases for larger inner cavities.  Consequently circumbinary disk lifetimes decrease to large $a$.  This effect is particularly notable in the strong wind case: photoevaporation rapidly erodes the disk, and the resulting lifetimes are a factor $\gtrsim10$ shorter than corresponding single-star disks.

%%%%%%%%%%%%%%%%%%%%%%%%%%
\section{Discussion}\label{sec:dis}
The 1-D models presented here are obviously simplified, and neglect several potentially important issues.  Non-axisymmetric features such as tidal accretion streams can only be modeled in an ad hoc manner, and I do not include the orbital modulation of the radiation fields that drive photoevaporation.  This simplified model also cannot account for eccentricity or inclination of the binary orbit, and neglects both the torques and shielding effects of circumstellar disks around each individual star.  More detailed two- and three-dimensional calculations are required to investigate these effects, but these simple models still provide an important initial insight into how young circumbinary disks evolve.

Perhaps the most intriguing result is that photoevaporative winds leave a potentially observable signature on the circumbinary disk distribution.  As seen in Figure \ref{fig:lifetimes}, all the models show an increase in disk lifetimes (by a factor $\lesssim 2.5$) for $a \lesssim 0.25 R_{\mathrm c}$.  Observations of circumbinary disk lifetimes can therefore potentially measure $R_{\mathrm c}$, which in turn would tell us the temperature (and origin) of the photoevaporative flow.  Current observations \citep[e.g.,][]{kraus12} do not yet provide a useful census of the young binary population at $\sim$AU separations, but future such surveys will offer an important diagnostic for protoplanetary disk dispersal.  I note also that my results are broadly consistent with observations of debris disks, which are much more frequent around close ($a < 3$AU) binaries than at wider separations \citep{trilling07}.

More generally, the fact that photoevaporation dominates circumbinary disk evolution for high wind rates suggests that individual binary systems can be used to test photoevaporative wind models directly.  \citet{kraus11,kraus12} found 10 Class II/III binaries in Taurus-Auriga with projected separations less than 50AU and detected disks (either circumstellar or circumbinary).  7 of these 10 objects have been detected in X-rays, with luminosities $L_{\mathrm X}\sim10^{29}$--$10^{30}$erg\,s$^{-1}$ \citep{xest}, but 9 of the 10 have estimated disk masses $\le 10^{-3}$\Msun\ \citep{aw05}.  The existence of so many low-mass circumbinary disks at an age of 1--2Myr suggests that they are not subject to strong photoevaporation, as the survival times for such disks are very short\footnote{Most interesting is perhaps DF Tau: this near-equal mass ($q=0.9$) binary has a projected separation of 10.6AU, a disk mass of $4\times10^{-4}$\Msun, and $L_{\mathrm X} \simeq 1\times10^{30}$erg\,s$^{-1}$ \citep{wk84,jkb97}.  This implies an X-ray photoevaporation rate of $7\times10^{-9}$\Msunyr\ \citep{owen11} and thus a disk lifetime of $\sim 5 \times 10^4$yr, which is uncomfortably short compared to the 1--2Myr age of the Taurus-Auriga association.  Note, however, that some Taurus binaries may be significantly younger than the single stars in the association \citep{kh09}.}$^,$\footnote{Also of interest is HD98800B, in the $\sim10$Myr-old TW Hya association \citep{furlan07,andrews10}.  This $a \sim 1$AU binary has a $3 \times 10^{-4}$\Msun\ circumbinary disk which extends from 3.5--15AU (where it is tidally truncated by HD98800A), and $L_{\mathrm X} = 1.4\times10^{29}$erg\,s$^{-1}$ \citep{kastner04}.  Again, the disk lifetime when subject to photoevaporation at the rate predicted by \citet{owen11}, $\sim 6\times10^5$yr, is much less than the system's age.}.  If we divide the observed disk masses by the stellar ages we can set a conservative upper limit to the photoevaporation rate of $\dot{M}_{\mathrm w}<10^{-9}$\Msunyr.  This is at least an order of magnitude smaller than predicted by models of X-ray photoevaporation, but roughly consistent with EUV photoevaporation models.  More detailed modeling and further observations of close binaries are still required, but the existence of large numbers of low-mass circumbinary disks seems to place a strong upper limit on the efficiency of disk photoevaporation.

Finally, it is interesting to consider the consequences of these results for planet formation.  We now know that giant planets around close binaries are relatively common \citep{doyle11,welsh12,orosz12a,orosz12b}, but little is known about their formation conditions.  My results suggest that circumbinary disks around close ($a \lesssim 1$AU) binaries are longer-lived than otherwise identical disks around single stars.  If disk lifetimes set a limit on the time-scale for planet formation, this suggests that planets can form readily in disks around close binaries.  However, photoevaporation causes a decline in circumbinary disk lifetimes with increasing binary separation, implying that circumbinary planets around wide ($a \gtrsim 10$AU) binaries should be rarer.

%%%%%%%%%%%%%%%%%%%%%%%%%%%%
\acknowledgments I thank Phil Armitage, Jim Pringle, Adam Kraus, Sean Andrews, Ilaria Pascucci, James Owen, Cathie Clarke and an anonymous referee for useful comments.  My research is supported by an STFC Advanced Fellowship (ST/G00711X/1).  Theoretical Astrophysics in Leicester is supported by an STFC Rolling Grant.

%%%%%%%%%%%%%%%%%%%%%%%%%%%%

%%%%%%%%%%%%%%%%%%%%%%%%%%%%%%%

\end{document}